\documentclass[aps,prb,twocolumn,groupedaddress]{revtex4}
\usepackage{graphicx} 

\begin{document}

\title{Shot Noise in Negative-Differential-Conductance Devices}

\author{W. Song}
\author{E. E. Mendez}
\author{V. Kuznetsov}
\author{B. Nielsen}
\affiliation{Department of Physics and Astronomy, State University of New York at Stony Brook, Stony Brook, NY 11794-3800}

\begin{abstract}
We have compared the shot-noise properties at T = 4.2 K of a double-barrier resonant-tunneling diode and a superlattice tunnel diode, both of which exhibit negative differential-conductance (NDC) in their current-voltage characteristics.  While the noise spectral density of the former device was greatly enhanced over the Poissonian value of $2eI$ in the NDC region, that of the latter device remained $2eI$.  This result implies that charge accumulation, not system instability, is responsible for shot-noise enhancement in NDC devices.
\end{abstract}

\maketitle

To gain a deep insight on the transport mechanisms of an electronic device it is sometimes essential to measure its shot noise. Then, the noise becomes the signal, paraphrasing Landauer.\cite{Lan1} The work described here illustrates dramatically this point, by comparing two types of semiconductor devices (a double-barrier resonant tunneling diode and a single-barrier tunneling diode with electrodes made out of doped superlattices) that exhibit similar non-linear current-voltage (I-V) characteristics and yet show very different shot-noise behavior.

In a double-barrier resonant tunneling diode, the current increases linearly with voltage once the first confined state in the well between the barriers is aligned in energy with the Fermi energy of the emitter electrode. As a consequence of parallel (to the heterostructure's interfaces) momentum conservation, the current drops abruptly when the confined state falls in energy below the conduction band of that electrode. It is this negative differential conductance (NDC) and the concomitant instability of the system that is attractive for high-frequency oscillators and even digital electronics.\cite{Miz}

The shot-noise properties of this device are well established both experimentally\cite{Zas,Bro,Ian,Kuz} and theoretically.\cite{Ian,Bla2,Bla1} The noise spectral density, S, is proportional to the current $I$, S = $2e$F$I$, where $e$ is the electronic charge and F is the so-called Fano factor. In the linear-current regime F is smaller than one, whereas in the NDC region it is larger than one. This suppression and enhancement of noise indicate, respectively, negative and positive correlations in the electronic motion. These departures from Poissonian noise (S = $2eI$) behavior are attributed to the accumulation of charge in the quantum well during the tunneling process,\cite{Ian,Bla2} but it has been pointed out by Blanter and B\"{u}ttiker that system instability, rather than charge accumulation, may be the ``necessary condition'' for shot noise enhancement.\cite{Bla1}
\begin{figure}
\includegraphics[width=81mm]{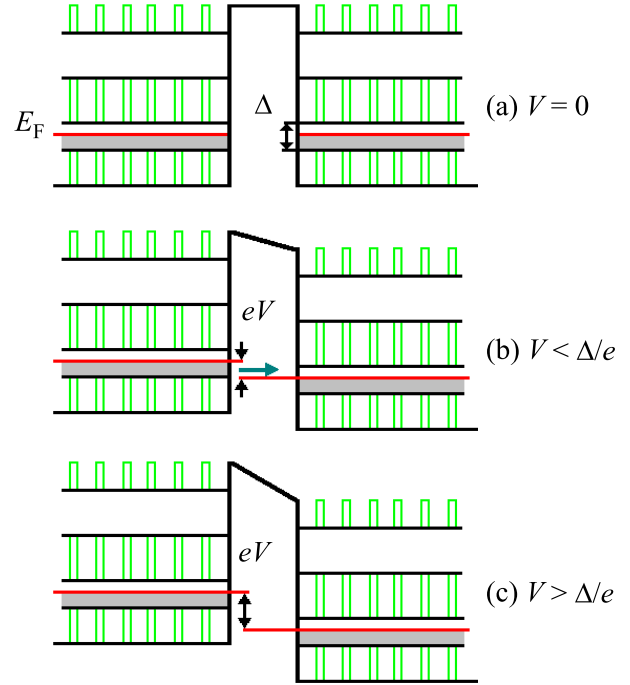}
\caption{\label{fBand} Energy profile of a superlattice tunnel diode, consisting of a wide potential barrier separating two identical, doped superlattices that act as electrodes.  In equilibrium, (a), the minibands of both superlattices are aligned in energy.  When a bias, V, is applied between the electrodes, (b), the bands are misaligned and electrons tunnel through the barrier. If the bias exceeds the value $\Delta/e$, (c), band misalignment is complete and the pseudo-gap between the first two minibands in the collector electrode prevents tunneling.}
\end{figure}

To explore that suggestion, we have measured the I-V characteristics and noise properties of a semiconductor diode that shows NDC behavior similar to that of a resonant-tunneling diode but in which charge accumulation does not occur. The device, which we call \emph{superlattice tunnel diode} and is sketched in Fig. \ref{fBand}, consists of a single potential barrier between two identical electrodes, each formed by a doped superlattice.\cite{Dav,Sle} For energies below that of the single-barrier height, each superlattice has two minibands separated by a one-dimensional energy gap (or pseudo-gap); only the ground-state miniband is (partially) occupied. On the application of a bias V, tunnel current flows between the first minibands of the electrodes, as long as eV is smaller than the miniband width, $\Delta$. But when the bias exceeds the critical voltage $\Delta/e$, at low temperature, the pseudo-gap between the collector's minibands ``blocks'' the tunneling electrons and the current drops abruptly. At even higher voltages, the current increases again when the collector's second miniband becomes aligned with the emitter's first miniband.
\begin{figure*}[thp!]
\includegraphics[width=150mm,clip=]{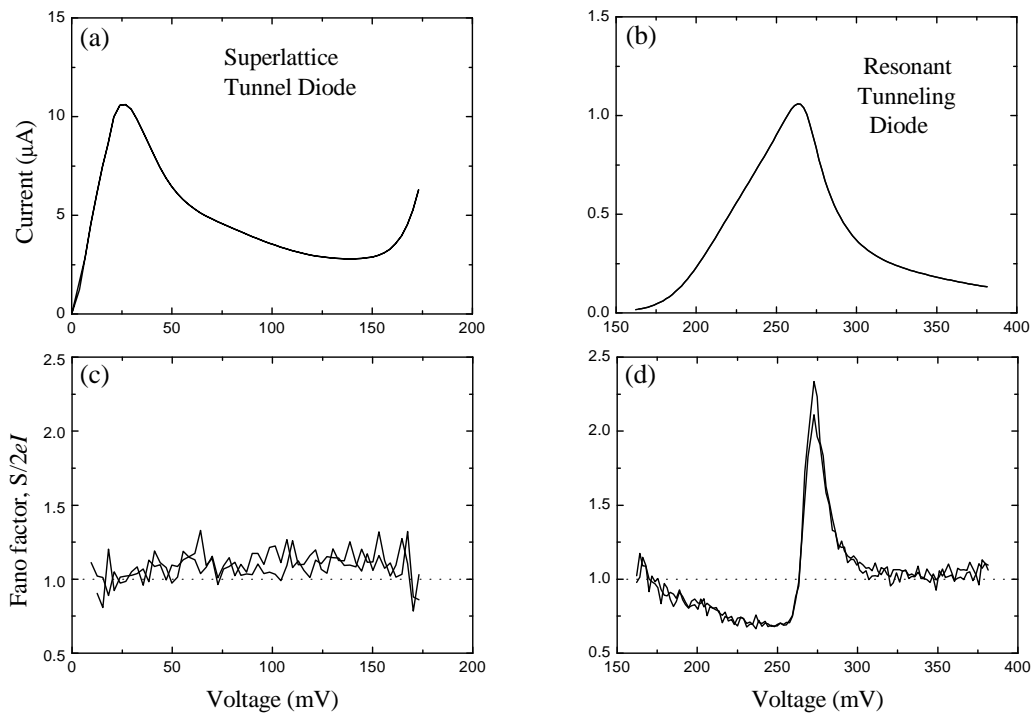}
\caption{\label{fSL-RTD} (a) Current-voltage characteristic, at T = 4.2 K, for a superlattice tunnel diode with an undoped 102 \AA  \ $\mathrm{Ga_{0.65}Al_{0.35}}$As barrier and 42 \AA\ - 23 \AA\ GaAs - $\mathrm{Ga_{0.65}Al_{0.35}}$As superlattices doped to 1$\times10^{18} \mathrm{cm}^{-3}$, as electrodes. (b) Current-voltage characteristic at 4.2 K for a double-barrier resonant tunneling diode with 100 \AA \ $\mathrm{Ga_{0.60}Al_{0.40}}$As barriers and a 40 \AA \ GaAs well. (c) Fano factor F (F = S/$2eI$, where S is the noise spectral density and $I$ is the current) for the diode of part (a). The value of F remains around 1, before and after the negative differential conductance region. (d) Fano factor for the double-barrier diode of part (b). In contrast with the result of (c), F falls below 1 before the peak-current voltage and is enhanced well above 1 in the negative differential conductance region. The double traces that appear in (c) and (d) correspond to measurements taken by sweeping the voltage up and down.}
\end{figure*}

The superlattice tunnel diode was implemented by preparing by molecular beam epitaxy a semiconductor heterostructure with an undoped 102 \AA  \  $\mathrm{Ga_{0.65}Al_{0.35}}$As barrier and a 50-period superlattice on each side of the barrier.  Each period was a 42 \AA\ - 23 \AA\ GaAs-$\mathrm{Ga_{0.65}Al_{0.35}}$As pair, doped uniformly with 1$\times10^{18} \mathrm{cm}^{-3}$ Si donors in one heterostructure (device A) and 5$\times10^{16} \mathrm{cm}^{-3}$ in a second, otherwise identical, structure (device B).  Two additional heterostructures were prepared as control samples.  In one, both electrodes were n$^{+}$ GaAs regions (device C) instead of superlattices.  The second control sample was a double-barrier diode with 100 \AA \ $\mathrm{Ga_{0.60}Al_{0.40}}$As barriers and a 40 \AA \ GaAs well, and with n-type GaAs electrodes (device D). In all cases, substrate and cap layers were heavily doped n-type GaAs, to which ohmic contacts were made by evaporating and annealing a AuGe-Ni alloy.

The I-V characteristic for device A at T = 4.2 K is shown in Fig. \ref{fSL-RTD}(a). Initially, the current increases almost linearly with increasing bias between the emitter and collector electrodes.  The quasi-linear dependence is followed by a current drop at around 25 mV, which continues up to 150 mV, at which point the current increases rapidly with bias. This behavior is consistent with the transport mechanism described above for a superlattice-tunel barrier diode. Sample B (not shown) exhibited a similar characteristic, although the peak-current and valley-current voltages were larger. 

For the material parameters of samples A and B, a simple Kronig-Penney calculation of the superlattice yields a 52 meV width for the ground-state miniband.\cite{Band} This number is considerably larger than the the 25 meV derived from the peak voltage in Fig. \ref{fSL-RTD}(a), although it agrees with the 50 meV obtained from the peak voltage for the opposite bias polarity (not shown). The asymmetry between the two bias directions is probably a consequence of an unintentional difference between the two superlattice electrodes. The larger voltages found for sample B result from light doping at the electrodes, where a significant portion of the applied voltage drops.  

Control sample C exhibited a non-linear I-V characteristic at 4.2 K but, as expected, no sign of NDC.  Finally, the double-barrier diode (sample D) had a sharp NDC region at 266 mV [see Fig. \ref{fSL-RTD}(b)], corresponding to the voltage at which the confined state in the well passes below the conduction band edge of the emitter.

Once the I-V characteristics of the various structures are well established, we can focus on their noise characteristics at 4.2 K.  In all cases, except at the smallest voltages, the noise was proportional to the current, an indication that the typical currents flowing through the devices were large enough to make thermal noise negligible.  At the same time, care was taken to restrict the currents (by a combinaton of design of the tunnel barriers and of device size) to low enough levels so that no hysteresis due to external-circuit instability was present in the NDC regions of the I-V characteristics.  

Figure \ref{fSL-RTD}(c), which shows the Fano factor of sample A as a function of voltage, captures the essence of this work: within experimental uncertainty, F is essentially one in the entire voltage range. As also observed in sample B, there is neither suppression nor enhancement before or after the NDC region. This is in marked contrast with the Fano factor we obtained for the double-barrier diode, sample D, shown in Fig. \ref{fSL-RTD}(d). Here F falls significantly below 1 in the quasi-linear current region and then shoots up sharply up to 2.3 at the NDC voltage, to fall back to 1 once the NDC region disappears.  

The sharp difference between the noise characteristics of the two types of NDC devices stems from the difference in their tunneling mechanisms. The partial suppresion of shot noise in the linear regime of the double-barrier diode has been explained by the charge correlation imposed by Pauli's exclusion principle, which prevents an electron from tunneling from the emitter into the quantum well unless the confined state is unoccupied. Noise suppression is thus a consequence of charge accumulation in the well. 

Theoretically, the Fano factor is F = 0.5 when the tunneling probabilities through the individual barriers are the same, and it approaches 1 when they are very different.\cite{Bla1} That in the case of Fig. \ref{fSL-RTD}(d) the minimum Fano factor is about 0.7 and not 0.5 implies an asymmetry between the two barriers, either intrinsic or induced by the applied voltage.

The noise enhancement in the NDC region of a double-barrier diode has been attributed to electrostatic-potential fluctuations when an electron tunnels into the well.  Once the confined state falls below the electrode's conduction-band edge and the tunneling density of states is very small, those fluctuations lead to positive feedback and charge correlations that increase shot noise.\cite{Ian,Kuz} A semiclassical, self-consistent calculation of the I-V and S-V characteristics using a linear approximation has shown that indeed the Fano factor diverges as the NDC voltage is approached.\cite{Bla2} 

In contrast, in the superlattice tunnel diode of Fig. \ref{fBand} there is no tunnel bottleneck and no charge accumulation. Electron tunneling is then an uncorrelated process and shot noise does not deviate from its Poissonian behavior, regardless of whether the diode is in the NDC regime or not. From the point of view of noise, the superlattice tunnel diode is not different from a conventional single-barrier heterostructure such as sample C, for which we found F=1 throughout the voltage range (V$\leq$140 mV).

In conclusion, we have compared two types of NDC devices and found experimentally that charge accumulation, not bistability, is what determines the non-Poissonian nature of shot noise. Early noise measurements on p-n tunnel diodes revealed full shot noise behavior,\cite{Tie} consistent with our conclusion, which we believe to be applicable to any situation in which NDC occurs.  

Simple shot-noise measurements like these presented here should allow to discern between competing transport mechanisms in which charge accumulation is an issue, even as subtle as in the case of tunneling through a single barrier into a quantum well.\cite{Mor} Such measurements might also help to differentiate between sequential and coherent tunneling in double-barrier heterostructures.  

We are grateful to M. Hong for providing some of the heterostructures used in this study, and to Ya. Blanter and M. B\"{u}ttiker for useful discussions. This work has been supported by the National Science Foundation and the Army Research Office.


\begin{thebibliography}{}
\bibitem{Band} The following values were used: barrier height, 310 meV; effective mass, 0.067 m$_0$, where m$_0$ is the free-electron mass.
\end{thebibliography}
\end{document}